\documentstyle[multicol,aps,prb,epsfig]{revtex}

\tighten
\begin{document}
\draft
\preprint{Submitted to Phys. Rev. B }

\title{ Dynamical Phase Diagrams for Moving Vortices Interacting with Periodic  Pinning}

\author{ Gilson Carneiro$^{*}$}
\address{Instituto de F\'{\i}sica\\Universidade Federal do Rio de Janeiro\\  
C.P. 68528\\ 21945-970, Rio de Janeiro-RJ \\ Brasil }
\date{\today}
\maketitle
\begin{abstract}

The dynamical phase diagrams for vortices in clean films, driven by an uniform force, and interacting with periodic pinning resulting from a  columnar-defect lattice are investigated by numerical simulations of a London model, and other considerations. Dynamical phases are identified according to the spatial symmetry, and  dynamical phase diagrams are studied as a function of the driving force magnitude and direction, and on the temperature, for typical vortex densities.  
The theoretical limit of infinite drive-force magnitudes is investigated first. In this limit vortex motion averages the periodic pinning potential in the direction of motion, and the dynamical phases reduce to thermal equilibrium ones for vortices interacting with the averaged pinning potential. For most directions of motion this potential is essentially constant. For a few particular ones it is a one-dimensional washboard, periodic  in the direction transverse to motion. The dynamical phase diagrams in this limit are obtained as a function of the direction of motion and temperature by equilibrium Monte Carlo simulations. They contain moving vortex lattices, commensurate or incommensurate with the washboards periodicities, at low temperatures,  moving smectics and  moving liquids at higher ones. Finite-drive dynamical phase diagrams are then obtained by numerical solution of Langevin's equations. It is found that each dynamical phase  originates from an infinite-drive limit phase with the same spatial symmetry that evolves continuously into finite-drive regions of the dynamical phase diagram. Dynamical transitions between moving commensurate or incommensurate lattices and moving liquids and between moving commensurate lattices and  moving smectics are observed.
Transverse pinning of  vortices at low temperatures is found in regions of the dynamical phase diagrams where the dynamical phases are moving commensurate lattices or  moving smectics.

\end{abstract}

\begin{multicols}{2}
\narrowtext
%\widetext

\section{introduction}
\label{sec.int}

A problem of  current interest is the study of moving vortices  interacting 
with arrays of pinning centers. Random arrays have received the most attention. However, periodic pinning arrays are also of  interest. One reason  is that they  provide examples of dynamical phases and phase transitions, a subject not yet fully understood, where theoretical predictions can be tested in superconducting films with artificial defect lattices\cite{ppa}, and in Josephson junction arrays (JJA)\cite{jja}.   

In the case of moving vortices interacting with random pinning arrays,  
the dynamics is simpler when the driving force is large, in which case 
the velocity of the center of mass (CM) of the vortex array is also large. 
This situation was first considered by 
Schmid and Hauger\cite{hdp} who  suggested that  fast moving vortices  average out the
pinning potential and, consequently, order in a triangular lattice at
low temperatures. In experiments \cite{rnde1} and numerical simulations\cite{rndn} 
nearly triangular lattices are indeed  observed at large  CM velocity. 
Koshlev and Vinokur \cite{kos} analyzed the effects of random pinning at large CM velocities on vortex displacement from the lattice equilibrium positions. They argued that it  is equivalent to a "shaking" temperature, inversely proportional to the CM velocity, that adds to the thermodynamic temperature, and  predicted that  dynamical  melting of the moving vortex lattice takes place when the CM velocity is such that the combined temperature equals the equilibrium melting temperature.
Recently,  Gianmarchi and Le Doussal \cite{gld} pointed out that this picture for the  large CM velocities behavior breaks down, because  the vortices order in a moving glass, rather than in a  moving triangular lattice. The reason is that averaging of the pinning potential by the fast moving vortices is only partial. A static random pinning potential remains acting on the  vortices in the frame of reference moving with the CM velocity (CM frame), and has non-trivial consequences on vortex order. These authors predict that the vortices move along static channels, on average parallel to the direction of drive, and are pinned on them with respect to  a transverse force at zero temperature. This  is referred to as transverse pinning. These predictions are in agreement with experimental and  numerical results \cite{rnde1}.

In the case of vortices interacting with periodic pinning, the dynamics  also simplifies in the limit of large driving forces. The question of what happens to  the dynamical phases in this limit was first examined in recent publications by the author \cite{gmc}. These papers consider pinning by a square lattice of columnar defects (CD), and vortex motion in the [0,1] (or [1,0]) direction. In Refs.\onlinecite{gmc} it is argued that  moving vortices average the periodic pinning potential only in the direction of motion. It is also argued  that the dynamical phases reduce to the equilibrium ones for vortices interacting with the periodic pinning potential averaged in the direction of motion, that is [0,1], which is a washboard, periodic  in the [1,0] direction. This was shown to have  non-trivial consequences for the dynamical phase diagram. The present paper studies further the  limit of large drives for the model introduced in Ref.\onlinecite{gmc}, allowing for arbitrary directions of vortex motion, and covering a wide temperature range. The dynamical phases in this limit are found to play a central role in determining the whole dynamical phase diagram.

Previous theoretical studies of vortices interacting with periodic pinning arrays are mostly numerical. Extensive studies of the 
zero-temperature vortex dynamics in systems with CD-lattices, carried out by several workers \cite{rndn}, find a  rather complex  behavior, with many dynamical phases 
and phase transitions. A simpler finite-temperature behavior is found in JJA and films.
Marconi and Domimguez \cite{dom1} carried out  numerical simulations of square JJA,  with the driving force along the [1,0] (or [0,1]) direction . They find three dynamical phases: two moving vortex lattices and a moving liquid. In all three, the CM velocity is along the direction of drive. The lattices have the same spatial order, but differ from each other by the response to a force transverse to the direction of motion. One lattice, named transversely pinned vortex lattice, is pinned  (transverse pinning), and the other, named floating solid, is unpinned. Dynamical  melting of the floating solid into the liquid, and dynamical depinning from the pinned lattice to the floating solid are reported. Further work \cite{dom2} predicts  transverse pinning only for driving forces along  [1,0]( or [0,1]), and  verify it  experimentally. 

Work on films was carried out  by Reichhardt and  Zim\'anyi\cite{rez} and by the author\cite{gmc}. Both investigate vortices interacting with a square CD-lattice  driven along the [1,0], or [0,1], directions, and consider the following vortex densities: $1$ vortex per CD in Ref.\onlinecite{rez};  $5/4$ and $2$ vortices per CD in Ref.\onlinecite{gmc}. They  find that the directions of vortex motion and drive coincide, and that the dynamical phase diagrams depend on the vortex density. For densities $1$ and $5/4$ vortices per CD the dynamical phase diagrams are similar, with three dynamical phases: a moving lattice, commensurate with the pinning potential periodicity along the direction transverse to motion, a  moving smectics and a moving liquid. Dynamical transitions between the commensurate lattice and the smectic and between the smectic and the liquid are found. In the commensurate lattice with $1$ vortex per CD tranverse pinning occurs \cite{rez}. For $2$ vortices per CD the only dynamical phases are an incommensurate lattice, essentially triangular, and a liquid, and dynamical melting between them takes place \cite{gmc}. 
The present paper shows how these particular  film results  fit into a dynamical phase diagram that  depends on the magnitude and direction of the driving force, and on the temperature. 

The approach adopted here is similar to that used in Ref.\onlinecite{gmc}. First the limit of high drives, where the dynamical phase diagrams are simpler, is  studied in detail. Only then the paper proceeds to investigate the intermediate and low drives behavior.  

The paper starts by considering Langevin's stochastic dynamical equations for a generic model for vortices interacting with periodic pinning at temperature $T$, and its behavior in the limit where the driving force magnitude approaches infinity. It is shown that in this limit these equations, written in the CM frame, describe vortices interacting between themselves and with a static effective pinning potential, equal to the average of the periodic pinning  potential in the direction of motion.  As a consequence, the dynamical phases reduce to the corresponding thermal equilibrium  ones, which are  hereafter referred to as  infinite-drive phases. These phases are exact asymptotic limits of dynamical phases occuring in the dynamical phase diagram.  The infinite-drive phases, being equilibrium phases, are distinguished from each other by the spatial order. Thus, dynamical phases with distinct spatial orders, one for each infinite-drive phase are expected to exist in finite-drive regions of the dynamical phase diagrams. The question is whether or not these dynamical phases exhaust the dynamical phase diagrams. The answer is found be yes for the model studied in this paper, as well as for those investigated by Marconi and Dominguez\cite{dom1} and by Reichhardt and  Zim\'anyi\cite{rez}.   

Next, the paper studies numerically the London model 
for vortices interacting with a square CD-lattice introduced in Ref.\onlinecite{gmc}. Two vortex densities, typical of the region where there are more vortices than CD ($B>B_{\phi}$), are considered: $2$ vortices per CD ($B=2B_{\phi}$) and $5/4$ vortices per CD
($B=1.25B_{\phi}$). The model dynamical phase diagrams for these two densities are obtained numerically. First, the model infinite-drive phases and phase diagrams are obtained as a function of the temperature and direction of drive by equilibrium Monte Carlo simulations of the London model for vortices interacting with the above mentioned effective pinning potential. Then dynamical simulations of  Langevin's equations describing the model dynamics are carried out. Based on the results of these simulations, the dynamical phases are identified and characterized, and the dynamical phase diagrams are obtained,  as a function of the driving force magnitude and direction and of the temperature. The main findings are. i) All dynamical phases have a corresponding infinite-drive phase with the same spatial symmetry. The portion of the dynamical phase diagram occupied by each dynamical phase is found to extend continuously into large driving-force magnitude regions, where the dynamical phase is essentially identical to the corresponding infinite-drive one.  
ii) Directions of drive and vortex motion are in general distinct, leading to anisotropic V-I characteristics. iii) Existence of two dynamical transitions. One, referred to as  dynamical melting,  between a moving incommensurate lattice and a moving  liquid. Another,  between a moving commensurate  lattice or a moving smectic and a moving liquid.
iv) Transverse pinning at low temperatures within regions of the dynamical phase diagram where the  spatial symmetry is that of a moving commensurate lattice  or a moving smectic, with vortex motion restricted to the [$\pm$ 1,1]  [0,1] and [1,0] CD-lattice directions.

\begin{center}
\begin{figure}
\epsfig{file=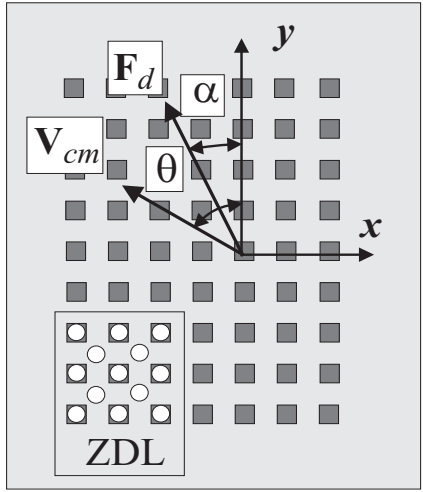,height=8cm,width=6.4cm,clip=}
\vspace{0.2cm}
\caption{ Columnar defect-lattice and definitions of  coordinate system, CM   velocity (${\bf V}_{cm}$) and driving force  ( ${\bf F}_{d}$) orientations. Inset: low-$T$  equilibrium vortex state at zero-drive for  $B=2B_{\phi}$ (ZDL).}     
\label{fig:fg1}
\end{figure}
\end{center}

A brief summary of the main numerical results is given next.
The CD-lattice  and the definitions of the coordinate system, CM velocity and driving force orientations are shown in Fig.\ \ref{fig:fg1}.
The model periodic pinning potential is shown in Fig.\ \ref{fig:fg2}.a. The  effective pinning potentials resulting from it are  washboards,  if motion is in the [0,1], [1,0] and [$\pm$1,1] CD-lattice directions, and essentially a constant otherwise. The [0,1] and [-1,1] washboards 
are shown in Fig.\ \ref{fig:fg2}.b and c. The [1,0] and [1,1] washboards follow from these by 
$90^o$ rotation. The  infinite-drive phases at low temperatures are found to be moving vortex lattices. For motion in the [0,1], [1,0] and [$\pm$1,1] directions these lattices can be commensurate or incommensurate with the washboard's periodicity, as shown in Fig.\ \ref{fig:fg3}. For motion in other directions  the moving  lattice is essentially triangular.   Dynamical phase diagrams for  driving forces oriented in the [0,1] and [-1,1] directions, in which case the directions of motion and drive coincide, are shown in Fig.\ \ref{fig:fg4}. The low-temperature dynamical phase diagrams as a function of the driving force magnitude and direction are shown in Fig.\ \ref{fig:fg5}. Within MIL  regions in Fig.\ \ref{fig:fg5} a moving commensurate lattice exists, with the vortices ordered in a nearly triangular lattice. Dynamical melting lines separate it from a moving liquid. In  the regions denoted MCL MCL45 and MSM in Fig.\ \ref{fig:fg5},  transverse pinning occurs, with vortex motion is restricted to the [-1,1] direction. The vortices order in moving commensurate lattices within MCL and MCL45, with  spatial symmetries equal to the infinite-drive phases with the same labels in Fig.\ \ref{fig:fg3}. Within MSM the spatial order is that of a moving smectic, similar to the MSM phase for $B=2B_{\phi}$ shown in Fig.\ \ref{fig:fg4}.  

This paper is organized as follows.  The infinite-drive limit is  considered in  Sec.\ \ref{sec.idl}. The London model  and the numerical methods used to study it are presented in  Sec.\ \ref{sec.num}. The simulation results are reported in Sec.\ \ref{sec.res}. Finally, in  Sec.\ \ref{sec.dis} a discussion of the paper main results and  comparison with earlier ones is presented, and the  conclusions are stated.

\section{infinite-drive limit}
\label{sec.idl}

Here the limit of infinite driving force magnitudes is discussed for a generic model for vortices interacting with a  CD-lattice.

The motion of $N_v$ two-dimensional vortices is assumed to be 
governed by Langevin equations for massless particles, which for the
$l$-th vortex reads\cite{ehb}, 
\begin{equation}
\eta \frac{d{\bf r}_l}{dt}= {\bf F}^{v-v}_l + {\bf F}^{v-cdl}_l
+  {\bf F}_{d} + {\bf \Gamma}_l\;, 
\label{eq.lan}  
\end{equation}
where $\eta$ is the friction coefficient,
\begin{equation}
{\bf F}^{v-v}_l=-\sum^{N_v}_{j\neq l=1}{\bf \nabla}_lU^{v-v}
({\bf r}_l-{\bf r}_j)\; ,
\label{eq.fvv}  
\end{equation}
is the force of interaction with other vortices, $U^{v-v}({\bf r})$ being the vortex-vortex interaction potential in two-dimensions,
\begin{equation}
{\bf F}^{v-cdl}_l=- {\bf \nabla}_l U^{v-cdl} ({\bf r}_l) \;, 
\label{eq.fvcdl} 
\end{equation}
is the force of interaction with the CD-lattice, $U^{v-cdl} ({\bf r})$, being the respective potential, given by
\begin{equation}
U^{v-cdl} ({\bf r})= \sum_{\bf R} U^{v-cd} ({\bf r}-{\bf R}) \;, 
\label{eq.uvcdl}  
\end{equation}
where ${\bf R}$ denotes the CD-lattice  positions and 
 $U^{v-cd}({\bf r})$
 is the interaction potential  between a vortex and a
single CD, ${\bf F}_{d}$ is the driving force, and
${\bf \Gamma}_l$ is the random force appropriate for  temperature $T$.
In terms of Fourier transforms $U^{v-cdl}_l$ can be written as 
\begin{equation}
U^{v-cdl} ({\bf r})=\sum_{\bf Q} U^{v-cd}({\bf Q}) e^{i{\bf Q}\cdot {\bf r}}\;,
\label{eq.uvcdlq}  
\end{equation}
where ${\bf Q}$ denotes the  CD-lattice
reciprocal lattice vectors, and $U^{v-cd}({\bf Q})$ is the Fourier
transform  of $U^{v-cd}({\bf r})$.  

It is convenient to consider vortex motion in the frame
moving with the CM. Let  ${\bf r}^{\prime}_l(t)= {\bf
r}_l(t)- {\bf V}_{cm}$t denote  the 
$l$-th vortex position in the CM frame, ${\bf V}_{cm}$
being the CM velocity, to be defined shortly. In the CM frame the vortex-CD lattice interaction, denoted here by ${\bf F}^{v-cdl}_l(t)$, depends explicitly on time, namely
\begin{equation}
{\bf F}^{v-cdl}_l(t)=\sum_{\bf 
Q}(-i{\bf Q}) U^{v-cd}({\bf Q}) e^{i{\bf Q}\cdot {\bf r}^{\prime}_l}
e^{i{\bf Q}\cdot {\bf V}_{cm}t} \;,
\label{eq.fvcdlt}  
\end{equation}
The equation of motion for the $l$-th vortex in the CM frame reads,  
\begin{equation}
\eta \frac{d{\bf r}^{\prime}_l}{dt}= {\bf F}^{v-v}_l + {\bf F}^{v-cdl}_l(t)
+  {\bf F}_{d} -\eta {\bf V}_{cm}+ {\bf \Gamma}_l\;, 
\label{eq.lanp}  
\end{equation}
The CM velocity is defined by the condition $ \sum^{N_v}_1 \langle d{\bf
r}^{\prime}_l(t)/dt)\rangle_t =0$ , where $\langle \rangle_t$
denotes average over the random force distribution and over time. It  follows from Eq.\ (\ref{eq.lanp}) that
\begin{equation}
 \eta{\bf V}_{cm}= {\bf F}_d 
+\frac{1}{N_v}\sum^{N_v}_{j=1} \langle {\bf F}^{v-cdl}_l(t)\rangle_t \;.
\label{eq.vcm}
\end{equation}

In the limit $F_{d}\rightarrow \infty$, it follows from  Eqs.\ (\ref{eq.fvcdlt}) and (\ref{eq.vcm}) that  $ {\bf V}_{cm}= {\bf F}_d/\eta $, so that $V_{cm}\rightarrow \infty$ also. In this case, all Fourier components in ${\bf
F}^{v-cdl}_l(t)$, Eq.\ (\ref{eq.fvcdlt}),  for which ${\bf Q}\cdot {\bf V}_{cm}\neq 0$ oscillate
fast,  having a negligible effect on the vortex trajectory \cite{hdp}, 
and ${\bf F}^{v-cdl}_l(t)$ reduces  to  the static
force obtained by summing 
the Fourier components in ${\bf F}^{(v-cdl)}_l(t)$ with ${\bf Q}\cdot
{\bf V}_{cm} = 0$, or ${\bf Q}\perp {\bf F}_d$, namely
\begin{equation}
{\bf F}^{v-cdl}_l(t)\rightarrow \sum_{{\bf 
Q}\perp {\bf F_d}}(-i{\bf Q}) U^{v-cd}({\bf Q}) e^{i{\bf Q}\cdot {\bf r}^{\prime}_l} \;
\label{eq.fid}  
\end{equation}
This force derives 
from the  static effective pinning potential 
\begin{equation}
U^{eff}_{\alpha}(r^{\prime}_{\perp})= \sum_{{\bf 
Q}\perp {\bf F_d}} U^{(v-cd)}({\bf Q}) e^{i Q r^{\prime}_{\perp}} \;,
\label{eq.ueff}  
\end{equation}
where $r^{\prime}_{\perp}$ is the CM frame
coordinate perpendicular to the direction of ${\bf F}_d$, and $\alpha$ the orientation of ${\bf F}_d$. This potential  is equal to the average of  $U^{v-cdl}$
in the direction of motion (and drive). By definition, $U^{eff}_{\alpha}(r^{\prime}_{\perp})$ is
one-dimensional and periodic in the direction perpendicular to ${\bf
F}_d$, if ${\bf F}_d$ is oriented along one of the  square CD-lattice
directions and a constant otherwise, since  no ${\bf Q}$ is perpendicular to
${\bf F}_d$. The equations of motion in the CM
frame, Eqs.\ (\ref{eq.lanp}),  reduce then to those describing  vortices interacting between themselves and  with  $U^{eff}_{\alpha}(r^{\prime}_{\perp})$ at temperature $T$.
Their long-time solutions are the corresponding equilibrium phases, called infinite-drive phases in this paper.

The infinite-drive limit establishes the exact asymptotic behavior $F_d\rightarrow \infty$ of the dynamical phase for  given ${\bf F}_d$ orientation and $T$.  
For finite $F_d$, the dynamical phase is expected to be close to the infinite-drive one with the same spatial symmetry for sufficiently large $F_d$. The condition for this to happen is that the vortex displacements in the CM frame caused by the  pinning force oscillations in time, Eq.\ (\ref{eq.lanp}), during a time interval of the order of one-half period, are small compared with the vortex mean separation $a_v$.  This means that for every ${\bf Q}$, such that  ${\bf Q}\cdot{\bf F}_{d} \neq 0$,  
 \begin{equation}
 Q\mid U^{(v-cd)}({\bf Q})\mid \frac{\pi}{{\bf Q}\cdot {\bf F}_d}\ll a_v \; .
\label{eq.idl}  
\end{equation}

\section{model and numerics}
\label{sec.num}

The model used in the numerical studies  carried out here  was introduced
in Ref.\onlinecite{gmc}. Vortices and CD are  placed on a square lattice (space
lattice), subjected to periodic boundary conditions 
with $N=256\times256$ square primitive unit cells of dimensions
$d\times d$. The CD-lattice is square,  oriented parallel to space lattice and commensurate with it. The CD-lattice, the coordinate system and the  angles defining the orientations of  ${\bf F}_d$  ($\alpha$) and ${\bf V}_{cm}$ ($\theta$) are shown in Fig.\ \ref{fig:fg1}.

The vortex has a core of linear dimension $d_v=4d$ ($d_v\sim 2\xi(0)$), so that each vortex occupies 16 space-lattice sites. The vortex-vortex interaction potential is  a screened  
Coulomb one \cite{min}, periodic in the  space lattice, and
characterized by the energy scale $J=(\phi^2_0d_v/32\pi^3\lambda^2)$, where $\lambda$ is the
penetration depth.
That is, $U^{v-v}({\bf r})$ is the lattice Fourier transform of 
\[U^{v-v}({\bf k})= 4\pi^2 J \frac{ e^{-\frac{\kappa^2}{\kappa^2_c}}}{ 
\kappa^2+\Lambda^{-2}}\; ,\]
where  $\kappa^2= 4\sin^2{(k_xd/2)}+4\sin^2{(k_yd/2)}$, $k_x$ and $k_y$ being the space lattice reciprocal lattice vectors components, 
$\kappa_c=2\sin{(\pi d/2d_v)}$ is the vortex core cutoff in k-space, and $\Lambda$ is
the screening length ($\Lambda>\lambda$).  
The square CD-lattice has $N_{cd}=8\times 8$ sites, lattice constant
$a_{cd}=32d$. The interaction
potential between a  vortex and a single CD is chosen with depth
$U^{v-cd}({\bf r}=0)=-J$, range $R_{cd}=12d$ and a spatial dependence
that gives square equipotentials and a pinning force of constant
modulus $F_p=J/R_{cd}$, as shown in Fig.\ \ref{fig:fg2}.a. 
The reciprocal CD-lattice vectors are
\[{\bf Q}=Q_1(n_1 \hat{\bf x} + n_2\hat{\bf y})\; ,\]
where $Q_1= 2\pi/a_{cd}$, and $n_1\;, n_2=-4,-3,...,3$. It is found that the lattice Fourier transforms $U^{v-cd}({\bf Q})$ are: $U^{v-cd}(0)=-0.19J$, 
$U^{v-cd}(\pm Q_1 \hat{\bf x})=U^{v-cd}(\pm Q_1 \hat{\bf y})=0.10J$, 
$U^{v-cd}(\pm Q_1[ \hat{\bf x} \pm \hat{\bf y}] )=-0.06$, and essentially negligible otherwise. 

The model has the 
square lattice symmetry, so that the dynamical phase
diagrams need only to be studied for ${\bf F}_{d}$ orientations
$0^o\leq \alpha \leq 45^o$. 

The dynamical phases spatial order is obtained by calculating  
the  time-averaged density-density correlation
function, $P({\bf r})$, which is proportional to the probability that a vortex 
is found at ${\bf r}$, given that there is one at ${\bf r}=0$, and its Fourier transform, the structure function, $S({\bf k})$. The motion of vortices is characterized by calculating the CM velocity, and  the time-averaged velocity of each vortex.

Two types of numerical studies of the model are carried out. Equilibrium Monte Carlo simulations  to obtain the infinite-drive phases, described in detail in Sec.\ \ref{sec.idph}. Numerical integration of Langevin's equations to obtain the dynamical phase diagram as a function of $F_d$, $\alpha$ and $T$ for two typical vortex densities: two 
vortices per CD  and five vortices per four CD. These corresponds to the  magnetic inductions  $B=2B_{\phi}$ and  $B=1.25B_{\phi}$, respectively, where $B_{\phi}=\phi_0/a^2_{cd}$ is the matching field, for which there is one vortex per CD. The numerical integration algorithm treats the vortex displacements as continuous during the run. In order to calculate the force acting on the vortices  at each time step, the vortex positions are reduced to the nearest space lattice ones. The accuracy of this approximation increases with the number of space lattice sites occupied by each vortex, equal to 16 here.  The numerical integrations are carried out at constant $B$, $\alpha$ and $T$, starting with a large $F_d$ and the vortex positions  initialized in a configuration appropriate for the infinite-drive phase.
Subsequent runs, each with a smaller $F_d$ than the previous one, are carried out  with the  vortex positions initialized in the last configuration obtained in the previous run,  until vortex motion can no longer be detected. In some cases, runs with $F_d$ retracing the decreasing sequences are carried out in order to check for hysteresis. 
Typical  runs consists of $\sim 0.4-1\times 10^6$ integration steps at intervals $10^{-2}\tau$, where $\tau=\eta d^2/J$ is the natural time scale. In order to average the  thermal force, the data  reported here is the  average over  $10$ repetitions of the same run, with different realizations of the random force.

\section{results}
\label{sec.res}

In this section  the simulation results for the dynamical phase diagrams are reported. First in  the  infinite-drive limit, then for finite drives. In the figures presented here driving force magnitudes are measured relative to the single CD pinning force $F_p$, Sec.\ \ref{sec.num}; temperatures relative to the  infinite-drive moving incommensurate lattice melting temperature $T_m$, Sec.\ \ref{sec.idph} and center of mass velocity components relative to the components of ${\bf V}_{d}\equiv{\bf F}_{d}/\eta$.
 
\subsection{Infinite-Drive Limit}

\label{sec.idph}

\begin{center}
\begin{figure}
\epsfig{file=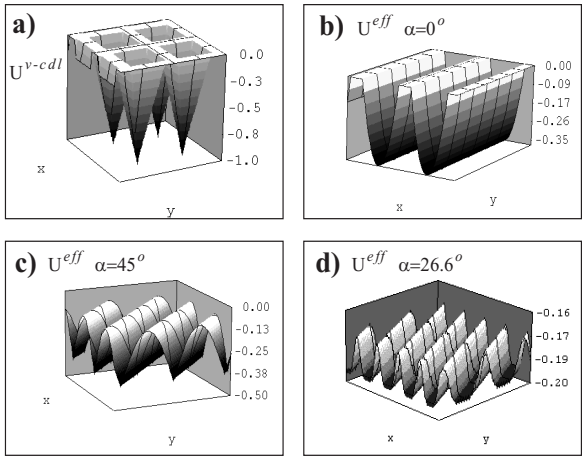,height=6.4cm,width=8.0cm,clip=}
\vspace{0.2cm}
\caption{ a) Columnar defect-lattice pinning potential. b-d) Effective pinning potentials for motion in: b)  [0,1] direction ($\alpha=0^o$), c) [-1,1] direction ($\alpha=45^o$), and 
d) [-1,2] direction ($\alpha=26.6^o$). Potentials in units of the single CD potential well depth ($J$).}     
\label{fig:fg2}
\end{figure}
\end{center}

\begin{center}
\begin{figure}
\epsfig{file=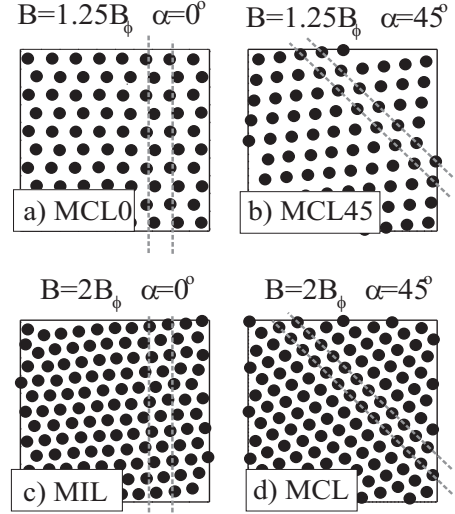,height=8cm,width=6.4cm,clip=}
\caption { Vortex positions in low-$T$ infinite-drive phases. Dashed gray lines indicate minima of $U^{eff}_{\alpha}$ shown in Figs.2.b and 2.c. Nomenclature: moving  commensurate lattices: a) MCL0, b) MCL45 and d) MCL. Moving incommensurate lattices: c) MIL } 
\label{fig:fg3}
\end{figure}
\end{center}

The  infinite-drive phase diagrams are obtained 
by  equilibrium Monte Carlo simulations of the lattice London  model
with the pinning  potential $U^{eff}_{\alpha}(r_{\perp})$.
It is found  that for ${\bf F}_d$ along [0,1] ($\alpha=0^o$)  and [-1,1] ($\alpha=45^o$) 
the $U^{eff}_{\alpha}(r_{\perp})$ are the washboards  shown in Figs.\ \ref{fig:fg2}b) and c). For ${\bf F}_d$ along other lattice directions, the $U^{eff}_{\alpha}(r_{\perp})$ are
found to be very shallow washboards, because the $U^{(v-cd)}({\bf Q})$ (Eq.\ (\ref{eq.ueff})) are very small ( Sec.\ \ref{sec.num}), and  are considered as constant 
potentials in this paper. For example, the $U^{eff}_{\alpha}(r_{\perp})$ for ${\bf F}_d$ along 
the [-1,2] direction ($\alpha=26.6^o$),  shown in Fig.\ \ref{fig:fg2}.d has a well depth more than one order of magnitude smaller than that  for  $\alpha=0^o$ and $45^o$ shown in Figs.\ \ref{fig:fg2}b) and c). For ${\bf F}_d$ oriented in non-lattice directions, $U^{eff}_{\alpha}(r_{\perp})$ is  a constant. The corresponding low-$T$ infinite-drive phases are as follows. For $\alpha=0^o$ and $45^o$, they are vortex lattices (VL) commensurate or incommensurate with the one-dimensional periodicity, depending on $B$. The VL for $\alpha=0^o$,   are obtained here as a function of $B$, for $B \geq B_{\phi}$. Their  density-density correlation functions, $P({\bf r})$,  are shown  in Fig.\ \ref{fig:fg6}. The  commensurate lattices consist of identical vortex chains within each washboard channel, with neighbor chains displaced with respect to each other by half a chain period. There is a single chain for $B_{\phi}\leq B \leq B_{1}$, and two chains for $B_{1}<B\leq B_{2}$. The values of $B_{1}$ and $B_{2}$ are found to be in the ranges $B_{1}<1.125B_{\phi}$ and  $1.125B_{\phi}<B_{2}<1.375B_{\phi}$. 
A commensurate-incommensurate  transition takes place for $B=B_{c-i}$ ($1.25B_{\phi}<B_{c-i}< 1.375B_{\phi}$). For $B>B_{c-i}$ the VL is incommensurate, and nearly triangular. For  $B=2B_{\phi}$ and  $B=1.25B_{\phi}$ the $P({\bf r})$ shown in Fig.\ \ref{fig:fg6} correspond, respectively,  to the vortex lattices labeled MIL and MCL0 in  Fig.\ \ref{fig:fg3}.
The effect of the one-dimensional periodic potential on the incommensurate lattice is to displace the vortices from the triangular lattice positions by a distance small compared to the lattice parameter\cite{cit}. It is found that, as expected, this effect is greater for $B$ in the vicinity of $B_{c-i}$.  This is  evidenced by the behavior of $P({\bf r})$ shown in 
Fig.\ \ref{fig:fg6}. For $B=1.375B_{\phi}$, $1.5B_{\phi}$, and $1.75B_{\phi}$, $P({\bf r})$ has smeared spots centered in a nearly triangular grid, with  smearing increasing with distance. For $B>2B_{\phi}$, $P({\bf r})$ has sharp spots, indicating  that the displacements  are negligible, and the  VL is sharply defined. 

The two $B$ values studied in this paper, $B=2B_{\phi}$ and  $B=1.25B_{\phi}$, are typical of parameter regions far from commensurate-incommensurate transitions, both for $\alpha=0^o$ and $\alpha=45^o$,  where the  VL are sharply defined. The infinite-drive phase diagrams in these cases are as follows. 
For $\alpha=0^o$ and $45^o$ the VL  are shown in Fig.\ \ref{fig:fg3}. These are referred to in this paper as moving incommensurate lattices  (MIL) and moving commensurate lattices, with distinct vortex configurations labeled MCL, MCL0 and MCL45, as shown in Fig.\ \ref{fig:fg3}. For $0^o<\alpha< 45^o$, the VL are incommensurate and nearly triangular. It is found that the incommensurate lattices for the same $B$, but different $\alpha$, cannot be distinguished from each other. Hereafter all incommensurate lattices are referred to as moving incommensurate lattices (MIL).

The $T$-dependence of the infinite-drive phases (at constant $\alpha$ is  found to be as follows:  the moving incommensurate lattices (MIL)  melts into a moving vortex liquid (MLQ) at $k_BT_{m}=0.09J$, for both $B$. For $B=2B_{\phi}$ the moving commensurate lattice (MCL) changes into a moving smectic (MSM) at $T_{mcl}/T_m=1.4$ and the moving smectic (MSM) changes  into a moving liquid (MLQ) at $T_{msm}/T_m=1.8$.  For 
$B=1.25B_{\phi}$ the moving commensurate lattices, MCL0 and MCL45,
 change into a moving smectic (MSM)  at $T_{mcl0}/T_m=1.2$  and $T_{mcl45}/T_m=1.7$, respectively.  

\subsection{Finite Drives}
\label{sec.fid}

First the results for  ${\bf F}_d$ oriented along the high symmetry directions [0,1] 
( $\alpha=0^o$), and [-1,1] ( $\alpha=45^o$) are presented. In these cases the directions of motion and drive coincide ($\theta=\alpha$) all the way down  to the  $F_d$ values where $ V_{cm}=0$. Next  results for $0^o <\alpha < 45^o$ are discussed.

\subsubsection{Drive along $\alpha=0^o$ and $\alpha=45^o$} 
\label{sec.a045}

\begin{center}
\begin{figure}
\epsfig{file=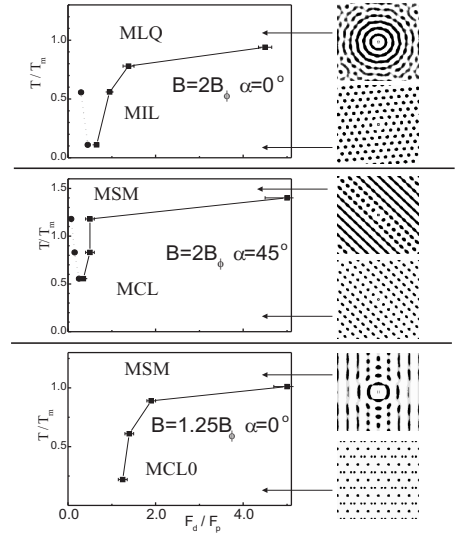,height=8.0cm,width=6.4cm,clip=}
\caption{ Dynamical phase diagrams  for drives along [0,1] ($\alpha=0^o$)  and [-1,1] ($\alpha=45^o$). Panels: density-density correlation functions $P({\bf r})$ at high drives (${\bf r}=0$ is at the panel's center). Nomenclature: MLQ=moving liquid, MSM=moving smectic. Others as in Fig.3. Dotted lines in the pannels for $B=2B_{\phi}$ indicate where vortex motion stops.} 
\label{fig:fg4}
\end{figure}
\end{center}

Dynamical phase diagrams ($F_d$ vs. $T$)  are  shown in Fig.\ \ref{fig:fg4}.   
For $B=2B_{\phi}$, $\alpha=0^o$, only two dynamical phases exist: a moving incommensurate lattice (MIL) and a moving liquid (MLQ), separated by a dynamical melting  line, as shown in Fig.\ \ref{fig:fg4}.  The dynamical phase diagrams for both $B$, $\alpha=45^o$, and  for $B=1.25B_{\phi}$, $\alpha=0^o$, are similar to one another, containing three phases: a moving commensurate lattice (MCL for $B=2B_{\phi}$, MCL0 and MCL45 for $B=1.25B_{\phi}$) a moving smectic (MSM) and a moving liquid. Only the dynamical phase diagrams for $B=2B_{\phi}$, $\alpha=45^o$, and $B=1.25B_{\phi}$, $\alpha=0^o$ are shown in Fig.\ \ref{fig:fg4}. That 
for $B=1.25B_{\phi}$, $\alpha=45^o$, is similar. In the temperature range covered by Fig.\ \ref{fig:fg4} the transitions between the moving smectics and moving liquids do not appear. The low $T$ properties of the $\alpha=0$ phase diagrams are studied in detail in Ref.\onlinecite{gmc}. In Fig.\ \ref{fig:fg4} typical $P({\bf r})$ at large $F_d/F_p$ for the above described phases are shown. It is found that the dynamical phases at  large $F_d / F_p$  essentially coincide with the corresponding infinite-drive ones. The dynamical phase diagrams  in Fig.\ \ref{fig:fg4} show clearly that each dynamical phase originates from 
the infinite-drive one with the same spatial symmetry that evolves  continuously into   finite drive regions.

For $B=2B_{\phi}$  the vortices freeze at low $T$ into the zero-drive lattice (ZDL in Fig.\ \ref{fig:fg1}) below the dotted lines in Fig.\ \ref{fig:fg4}. In this case the phase diagrams show little irreversibility. That is, if $F_d$ is increased staring from the zero drive lattice essentially the same results are obtained.
For $B=1.25B_{\phi}$ the vortices  do not freeze at low $T$ into the equilibrium  zero drive lattice, but remain in a metastable state, similar to the moving smectic, down to very small $F_d$.  The zero-drive lattice in this case is found to be a complex VL,  commensurate with the full CD-lattice potential, but with many vortices per unit cell. However, if $F_d$ is increased starting from the metastable state, little hysteresis is found.
  
The  transition lines show in Fig.\ \ref{fig:fg4} indicate the existence of two regimes for the dynamical transitions. One, for  large $F_d$, where the transition temperature depends  weakly on $F_d$, being close to its equilibrium values in the infinite-drive limit, showing that  the transitions are driven by thermal fluctuations.  The other, at small $F_d$, where the transition force depends weakly  on $T$, being close to the limit value of $F_d$ as $T\rightarrow 0$, indicating that the transitions are driven by vortex vibrations, caused by  motion in the periodic pinning potential. The latter mechanism is similar to vortex 'shaking', proposed in Ref.\onlinecite{kos} for random pinning, but a 'shaking' temperature cannot be defined. 

As discussed next, the commensurate VL ( MCL, MCL0 and MCL45) show transverse pinning at low $T$. This may be expected because for $\alpha=0^o$ and $\alpha=45^o$, ${\bf V}_{cm}$ remains parallel to ${\bf F}_{d}$. In all these cases, according to Eq.\ (\ref{eq.fvcdlt}), the effective pinning potential ($U^{eff}_{\alpha}(r_{\perp})$) acts on the moving vortices for all $F_d$, not only in the $F_d \rightarrow \infty$ limit. The commensurate phases are pinned by $U^{eff}_{\alpha}(r_{\perp})$ with respect to a small  force along the direction defined by $r_{\perp}$. 

\subsubsection{Drives along $0^o<\alpha< 45^o$}
\label{sec.aot}

The dynamical phases found in this range of driving-force orientations have the same spatial symmetries as those described in Secs.\ \ref{sec.idph} and  \ref{sec.a045} and are referred to in what follows by the same nomenclature. 

\begin{center}
\begin{figure}
\epsfig{file=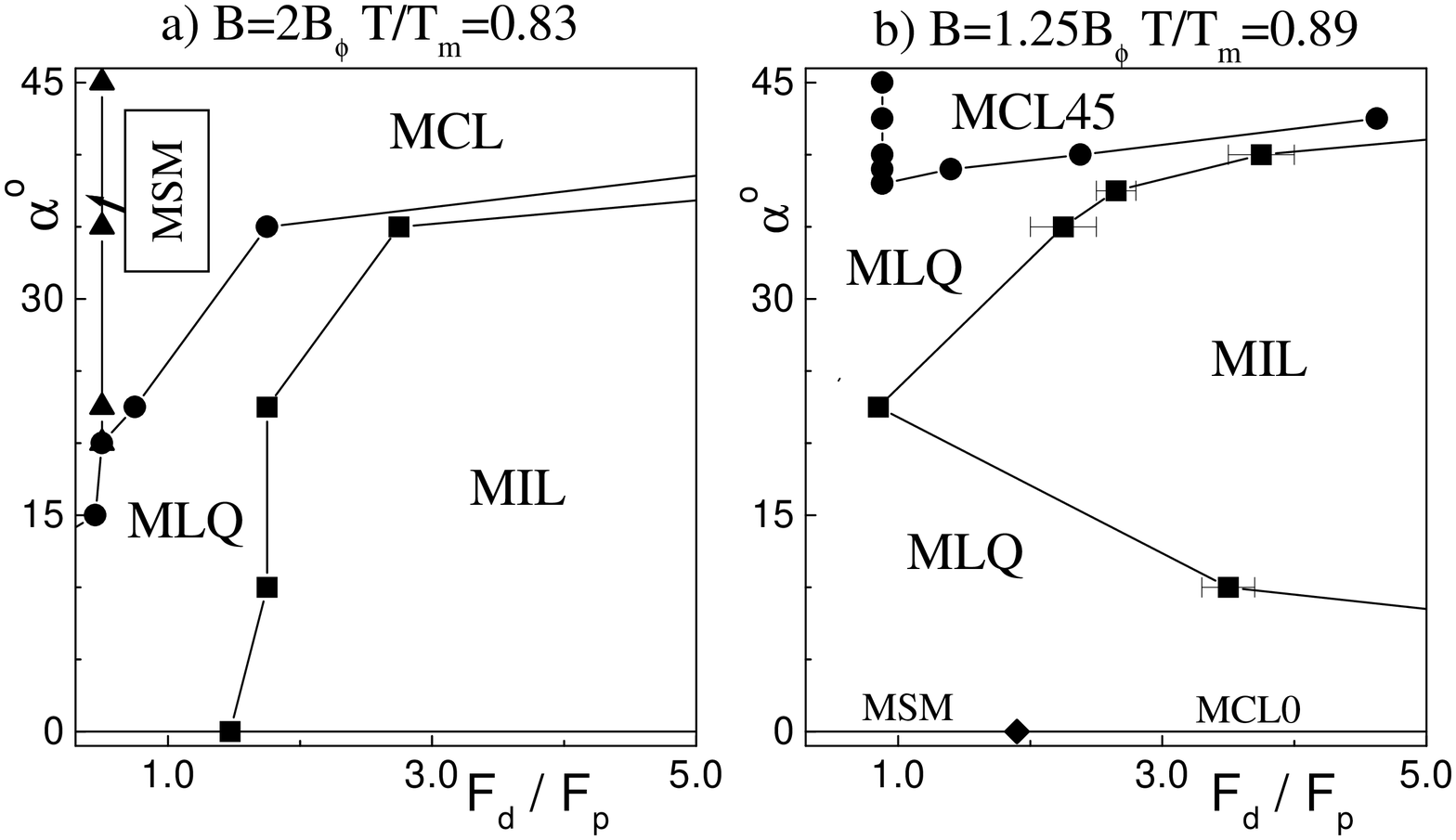,height=6.4cm,width=8.0cm,clip=}
\caption{ Dynamical phase diagrams for $T=$ constant. Dynamical phases 
named  as in Figs. 3 and 4.}    
\label{fig:fg5}
\end{figure}
\end{center}

The  $F_d$ vs. $\alpha$ dynamical phase diagrams at low-$T$ are shown in Fig.\ \ref{fig:fg5}. 
Both have two dynamical transition lines. One, referred to here as dynamical melting line,  separating a moving incommensurate lattice (MIL) and a moving liquid (MLQ). Another separating a moving commensurate lattice (MCL for $B=2B_{\phi}$ and MCL45 for $B=1.25B_{\phi}$) or a moving smectic (MSM) for $B=2B_{\phi}$ and a moving liquid. Within the  moving commensurate lattice and moving smectic regions transverse pinning occurs, with the vortices moving along the [-1,1] direction.  The transition line in the $B=2B_{\phi}$  dynamical phase diagram from the moving  commensurate lattice (MCL) and the moving smectic (MSM) occurs with the vortices moving in the [-1,1] direction , and is essentially identical to that for $\alpha=45^o$ discussed in  Sec.\ \ref{sec.a045}

For clarity of presentation, the dynamical phase diagrams of Fig.\ \ref{fig:fg5} do not show the $F_d$ vs. $\alpha$ line representing the cessation of vortex motion at small $F_d$. It is found that  the behavior is similar to that for $\alpha =0^o$ and $45^o$ discussed in  Sec.\ \ref{sec.a045}. For $B=2B_{\phi}$ the vortices freeze in the zero-drive lattice (ZDL)  shown in Fig.\ \ref{fig:fg1} for all $\alpha$. For $B=1.25B_{\phi}$ freezing of the vortices occurs in a metastable state which change somewhat with $\alpha$.

\begin{center}
\begin{figure}
\epsfig{file=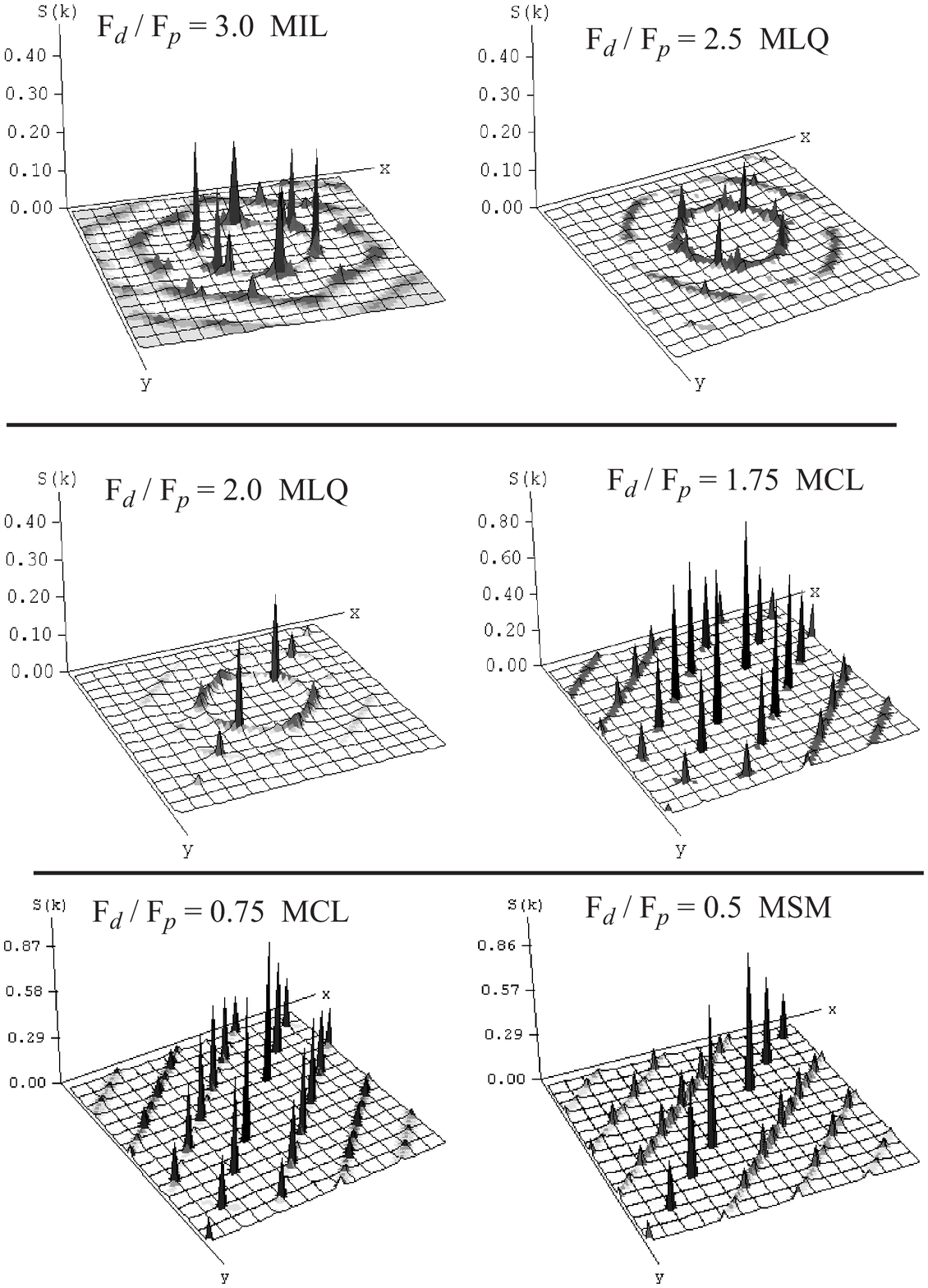,height=8.0cm,width=6.4cm,clip=}
\caption{ Dependence of the structure function, $S({\bf k})$, for  $B=2B_{\phi}$ and $T/T_m=0.83$ on driving force, $F_p$, along the $\alpha=35^o$ line in Fig.5.a. Spatial symmetries named as in Fig.3 and 4. } 
\label{fig:fg7}
\end{figure}
\end{center}

\begin{center}
\begin{figure}
\epsfig{file=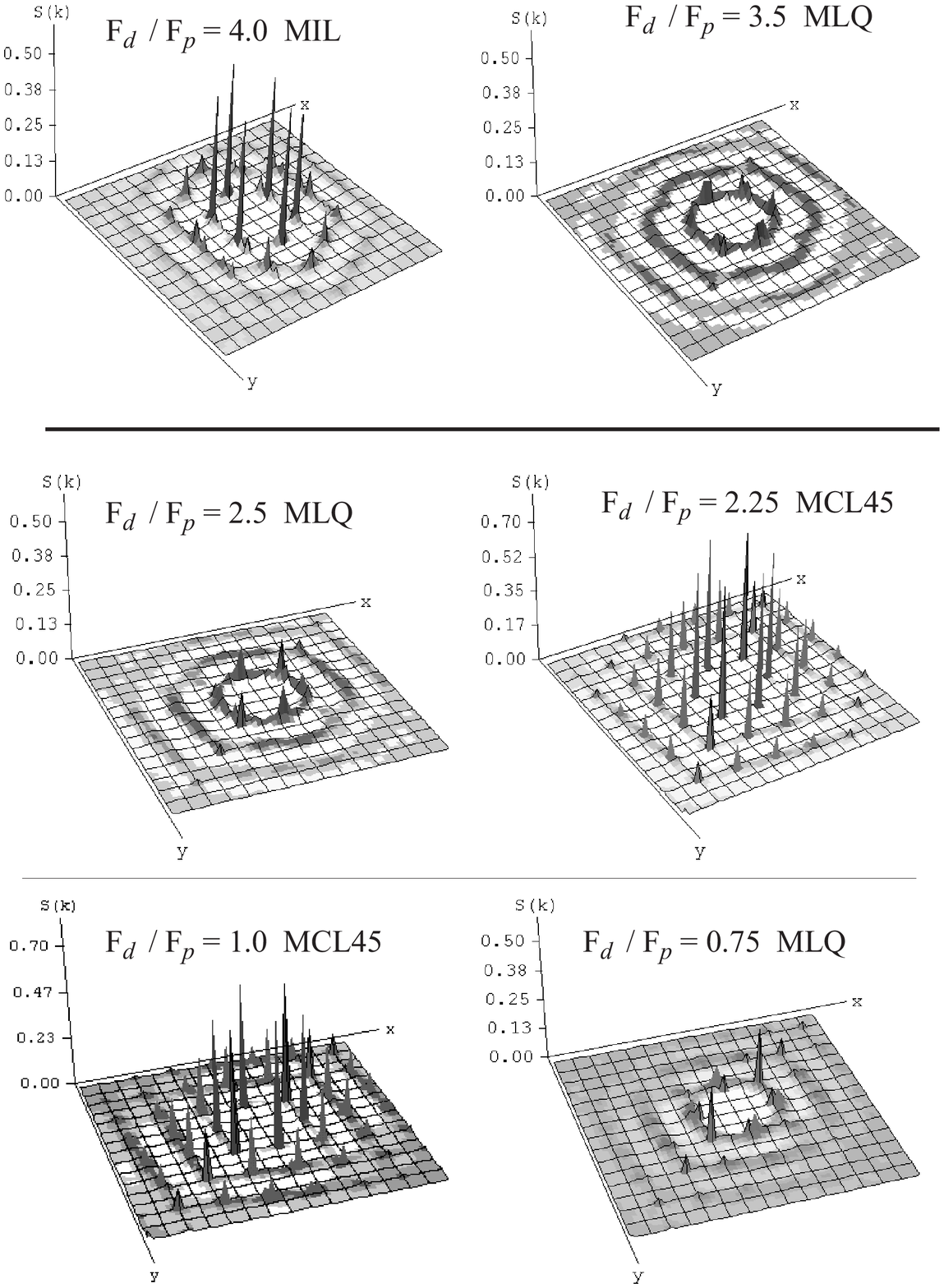,height=8.0cm,width=6.4cm,clip=}
\caption{ Dependence of the structure function, $S({\bf k})$, for $B=1.25B_{\phi}$ and 
$T/T_m=0.89$ on  driving force, $F_P$, along the $\alpha=40^o$ line in Fig.5.b. Spatial symmetries named as in Fig.3 and 4. } 
\label{fig:fg8}
\end{figure}
\end{center}

\begin{center}
\begin{figure}
\epsfig{file=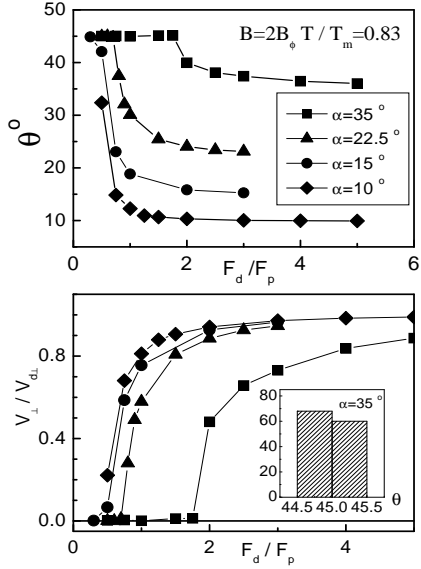,height=8.0cm,width=6.4cm,clip=}
\caption{ Center of mass velocity direction ($\theta$) and component perpendicular to the [-1,1] direction , $V_{\perp}$, vs. driving force magnitude along  lines of constant $\alpha$ in Fig.5.a.  Inset: histogram for the distribution of  vortices direction of motion ($\theta$) for $\alpha=35^o$ and $F_d/F_p=1.5$. }
\label{fig:fg9}
\end{figure}
\end{center}

\begin{center}
\begin{figure}
\epsfig{file=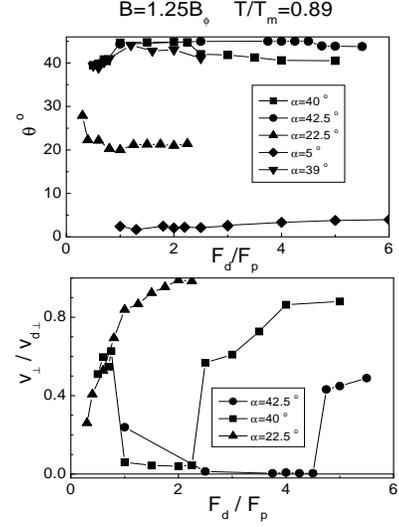,height=8.0cm,width=6.4cm,clip=}
\caption{  Center of mass velocity direction ($\theta$) and component perpendicular to the [-1,1] direction ,$V_{\perp}$, vs. driving force magnitude along  lines of constant $\alpha$ in Fig.5.b. }  
\label{fig:fg10}
\end{figure}
\end{center}

The evidence leading to the construction of the dynamical phase diagrams  in Fig.\ \ref{fig:fg5} is summarized in Figs.\ \ref{fig:fg7}, \ref{fig:fg8}, \ref{fig:fg9}, and \ref{fig:fg10}, where the  
dependencies of the spatial order and ${\bf V}_{cm}$  on $F_d$, along lines of constant $\alpha$ in Figs.\ \ref{fig:fg5}.a and b,  are shown.  The evolution with $F_d$ of  $S({\bf k})$ is shown in 
Figs.\ \ref{fig:fg7} and \ref{fig:fg8} along the $\alpha=35^o$ line  for $B=2B_{\phi}$, and along the  $\alpha=40^o$ line for $B=1.25B_{\phi}$. The changes occurring in  $S({\bf k})$ shown in these figures when $F_d$ crosses one of the transition lines  are typical of those occuring across the same line at different  $\alpha$. In Figs.\ \ref{fig:fg9} and \ref{fig:fg10} the direction of motion ($\theta$) vs. $F_d$ curves  are shown for  various $\alpha$. Also shown in is the component of ${\bf V}_{cm}$ perpendicular to the [-1,1] direction, denoted $V_{\perp}$, as a function of $F_d$. This quantity is proportional to the  voltage difference between contacts placed perpendicular to the [-1,1] direction. In Figs.\ \ref{fig:fg9} and \ref{fig:fg10}. $V_{\perp}$ is measured relative to the  component of ${\bf V}_{d}\equiv{\bf F}_{d}/\eta$ perpendicular to the [-1,1] direction, denoted by $V_{d\perp}$. 

The detailed properties of the dynamical phase diagrams shown in   Fig.\ \ref{fig:fg5} are as follows. 

\noindent i) $Dynamical Melting.$ The dynamical melting lines extend from $\alpha = 0^o$ to $\alpha= 45^o$. For $B=2B_{\phi}$ it touches the $\alpha=0^o$-axis
at the $F_d$ value where the dynamical melting takes place for $\alpha=0^o$   and $T/T_m=0.83$ (Fig.\ \ref{fig:fg4}). It does not touch the $\alpha=45^o$ axis for both $B$ and the  $\alpha=0^o$ axis for $B=1.25B_{\phi}$, because the dynamical phases in these axes are not moving incommensurate lattices, as discussed in Sec.\ \ref{sec.idph}. These result of 
(Fig.\ \ref{fig:fg5})show that, for both $B$, the moving incommensurate  and commensurate lattices are separated by a moving liquid, at least up to the highest $F_d$ studied here. However, in the infinite-drive phase diagram there is no moving liquid at these temperatures. It is unclear  how the disappearance of the moving liquid  as $F_d\rightarrow \infty$  takes place.  

The $\theta$ vs. $F_d$ curves in   Figs.\ \ref{fig:fg9} and \ref{fig:fg10} show that in the moving incommensurate lattice (MIL) the directions of drive and vortex motion are approximately equal ($\theta \simeq \alpha$), and that  the melting transition is reflected in this curve by a change in slope. For  $B=2B_{\phi}$ this change is  sharp, with $\theta$ increasing rapidly towards $\theta=45^o$ as $F_d$ decreases.  For $B=1.25B_{\phi}$ the same is true,  as long as $\alpha\gtrsim 20^o$, while for smaller $\alpha$ the moving liquid $\theta$ slowly approaches $\theta =0^o$ as $F_d$ decreases. 

It is found that the moving liquids are anisotropic, particularly when $\theta$ is close to $\theta=45^o$. This is evidenced by the  $S({\bf k})$ panels in Fig.\ \ref{fig:fg7}  for $F_d/F_p=2.0$, and in Fig.\ \ref{fig:fg8}  for $F_d/F_p=2.5$ and $F_d/F_p=0.75$, where  a pair of  peaks  at ${\bf k}=\pm Q_1(\hat{\bf x} +\hat{\bf y})$ occurs. These peaks result from a modulation in the vortex density  along [1,1], with wavelength $\sqrt{2}a_{cd}$,  due to  vortex motion being predominantly along the [-1,1] direction ($\theta \sim 45^o$, see Figs.\ \ref{fig:fg9} and \ref{fig:fg10}), which  favor concentration of vortices around lines in the 
[-1,1] direction connecting the CD-lattice potential minima (Fig.\ \ref{fig:fg2}.a).  

ii) $Transverse Pinning$. The  results shown in Figs.\ \ref{fig:fg9} and \ref{fig:fg10} indicate  that in the regions where the spatial order is that of a moving commensurate lattice, or a moving smectic for $B=2B_{\phi}$,   transverse pinning occurs, and  vortex motion is restricted to the 
[-1,1] direction. This  is seen in Figs.\ \ref{fig:fg9} and \ref{fig:fg10}, where  $V_{\perp}=0$ and $\theta=45^o$  in 
the $F_d$ range of moving commensurate lattices or smectics, and the distribution of individual vortices direction of motion is sharply peaked around $\theta=45^o$ (Fig.\ \ref{fig:fg9}, inset). In the literature, tranverse pinning is usually discussed in terms of the  ${\bf F}_{d}$ component perpendicular to the direction of motion needed to depin the vortices. For the results   shown in Fig.\ \ref{fig:fg5} this quantity, denoted here by $F_{\perp c}$, can be obtained  from the transition lines between the moving commensurate lattices, or smectics, and the moving liquid by $F_{\perp c}=F_d\sin{(45^o -\alpha)}$, where  $F_d$ and $\alpha$ are taken  at these lines.  
These results also show that the moving commensurate and smectic regions  evolve continuously from the corresponding infinite-drive phases. As pointed out in Sec.\ \ref{sec.idph}, transverse pinning can be anticipated when the infinite-drive phase is a commensurate lattice. In the cases where this  lattice is a single- chain one, $\alpha=45^o$ both $B$ (Fig.\ \ref{fig:fg3}), the depinning force can be estimated as being that for a single vortex, which is equal to the maximum force exerted by the $\alpha=45^o$ washboard ( Fig.\ \ref{fig:fg2}.c).  The  value of the maximum force is found to agree, within the simulation errors, with $F_{\perp c}$ obtained  from  the data shown in  Fig.\ \ref{fig:fg5} at large  $F_d$.  In the case $B=1.25B_{\phi}$, it is expected that a moving commensurate lattice region with spatial symmetry MCL0   occurs near  $\alpha=0^o$. This region is not observed  for $T/T_m=0.89$,  used  in Fig.\ \ref{fig:fg5}, but lower $T$ results indicate that it exists,  and is restricted to the immediate vicinity of  $\alpha=0^o$. This may result from a very small transverse depinning force  for the  MCL0  moving commensurate lattice, due to its two-chain structure. This makes it difficult to obtain accurate results for this region, and it will not be discussed further in this paper.

\begin{center}
\begin{figure}
\epsfig{file=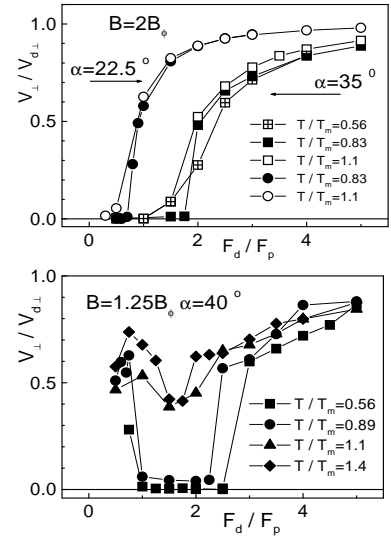,height=8.0cm,width=6.4cm,clip=}
\caption{ Component of CM velocity perpendicular to [-1,1], $V_{\perp}$, vs.  driving force magnitude along lines of constant $\alpha$  and at constant temperatures. }
\label{fig:fg11}
\end{figure}
\end{center}

The temperature dependence of the dynamical phase diagrams is as follows.
For temperatures lower than  those of Fig.\ \ref{fig:fg5} the dynamical phase diagrams are similar. The dynamical melting lines are shifted to lower $F_d$, similarly to what happens for $B=2B_{\phi}$ and $\alpha=0^o$ (Fig.\ \ref{fig:fg4} ). Transverse pinning in the moving commensurate lattices also occurs, as evidenced by the $V_{\perp}$ vs. $F_d$ curves shown in Fig.\ \ref{fig:fg11}. For higher temperatures several changes are observed.  The dynamical melting lines  also exist, shifted to larger $F_d$, provided  $T$ does not exceed the  infinite-drive melting  temperature $T_m$. When $T>T_m$ the moving incommensurate lattices do not exist, only moving liquids.

As well known\cite{gld}, transverse pinning in the moving commensurate lattices or smectics only exists, strictly speaking, at $T=0$. At finite $T$ thermal excitation  of lattice defects lead to flux motion away from the [-1,1] direction, giving rise to a finite $V_{\perp}$. The regions of transverse pinning found in the present simulations occur because  the  thermally excited $V_{\perp}$ is smaller than the simulation resolution. In experiments, where the resolution is also finite, these regions may also be present. The $T$ dependence of the $V_{\perp}$ vs $F_d$ curves are shown in Fig.\ \ref{fig:fg11}. For $B=2B_{\phi}$ transverse pinning is found in the moving commensurate lattice and moving smectics at temperatures above $T_m$, whereas for $B=1.25B_{\phi}$ it disappears at lower temperatures.

\section{discussion}
\label{sec.dis}

The  results described in Sec.\ \ref{sec.res} reveal the important role played by the infinite-drive phases. They  show that all  dynamical phases originate from infinite-drive phases with the same spatial symmetry that evolve continuously into regions of finite drives.  
 
\begin{center}
\begin{figure}
\epsfig{file=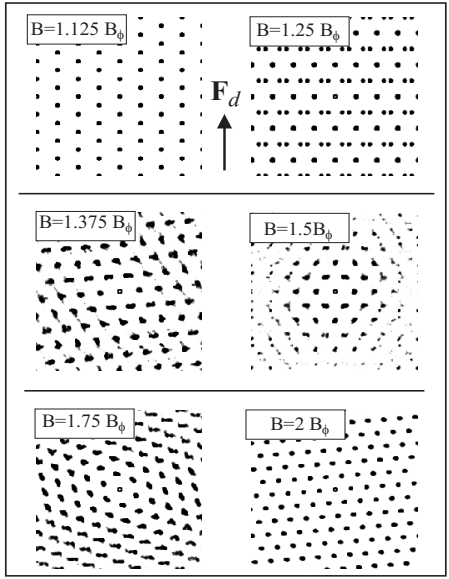,height=8.0cm,width=6.4cm,clip=}
\caption{ Density-density correlation functions, $P({\bf r})$ (${\bf r}=0$ is at the panel's center), for low-$T$ infinite-drive phases for motion in the [0,1] ($\alpha=0^o$) direction as a function of $B$.} 
\label{fig:fg6}
\end{figure}
\end{center}

The results obtained by Reichhardt and  Zim\'anyi\cite{rez} for vortices in films interacting with a square CD-lattice, driven in the [1,0] direction,  at $B=B_{\phi}$, show a similar  behavior. The infinite-drive phase diagram  was not considered in this paper, but can be identified with the reported dynamical phase diagrams at high drives.
It consists, in the terminology adopted here, of the following phases.  A moving commensurate lattice at low $T$, with single-chain structure similar to that shown in Fig.\ \ref{fig:fg6} for $B=1.125B_{\phi}$ rotated by $90^o$,  a moving smectic at intermediate $T$, and a moving liquid at high $T$. The dynamical phase diagram  reported in Ref.\onlinecite{rez} shows that these three phases exist in  continuous regions of the driving force vs. $T$ phase diagrams that extend from high to low drives.  This phase diagram  is similar to those found here for $B=1.25B_{\phi}$,  $\alpha=0^o$ and  $45^o$, and for $B=2B_{\phi}$  $\alpha=45^o$ ( Sec.\ \ref{sec.a045}).

For vortices in JJA, driven in the [1,0] direction, the dynamical phase diagram reported by Marconi and Dominguez\cite{dom1} also show dynamical phases  existing in  continuous regions of the driving force vs. $T$ plane  that extend from high to low drives. However it is not possible to identify the infinite-drive phases from the data reported in the paper.

Some studies of vortices interacting with CD-lattices at $T=0$ \cite{rndn}  find dynamical phases and phase transitions similar to the ones reported in this paper and in Ref.\onlinecite{rez}. Examples are tranverse pinning, moving commensurate and incommensurate lattices and dynamical transitions between them. However, these studies probe dynamical  behavior different  from the finite-temperature ones discussed here. At finite temperatures, the vortices relax after sufficient long times to a steady state, identified as the dynamical phase, that is unique as far as the probability distribution is concerned \cite{fkpe}. Accordingly,  dynamical phase changes caused by varying the  driving force, the temperature, or both, are reversible. This is not necessarily the case at $T=0$. However, in numerical studies, irreversibility appears  due to insufficient run times to reach the steady state, particularly when relaxation times are very large. This occurs in 
several circumstances, such as low temperatures, or vicinity of dynamical phase changes.
In the  simulations reported in this paper  low temperatures are avoided. Some simulation runs where performed cycling  the driving force magnitude from a large value to a small one and back. Small hysteresis is found near dynamical phase transitions.

In conclusion, the approach adopted in this paper to construct the dynamical phase diagrams, 
starting from the infinite-drive ones,  provides a simple method, based on equilibrium statistical mechanics, to identify  dynamical phases spatial order, and to predict dynamical phase transitions. It is expected that this method is  applicable to a large class of periodic pinning potentials and vortex densities. The reasons are that the effective pinning potentials resulting from averaging  physically reasonable  periodic pinning potentials in the direction of motion are, as those discussed in Sec.\ \ref{sec.idph}, essentially constant for most directions, and one-dimensional and periodic  for a few particular ones. This predicts the existence, for $B\geq B_{\phi}$, of moving lattices, commensurate or incommensurate with the effective pinning potential periodicity, moving smectics and moving liquids, similar to the ones reported here,  and of dynamical phase transitions  between then.  For $B<B_{\phi}$ the moving lattices may be different, particularly at low vortex densities, where the  commensurate infinite-drive phases are not expected to retain the simple chain structure found here for the $B\geq B_{\phi}$ ones. However, the close relationship between the dynamical phase diagrams and the infinite-drive ones are still valid, as evidenced by the JJA results of Ref.\onlinecite{dom1}. The dynamical phases obtained by this method do not, in general, exhaust the dynamical phase diagram. One known type of dynamical phase that has no corresponding infinite-drive phase is that in which some vortices are pinned by the periodic potential and others are moving. These are found at low drives in some $T=0$ simulations for $B>B_{\phi}$ \cite{rndn}.  No such phases are found in the present simulations nor in Refs.\onlinecite{rndn,dom1}. The details of how the dynamical phases and dynamical phase transitions predicted by the method proposed here fit into the  dynamical phase diagram for each particular model and vortex density depends in a complicated way on the model parameters, and has to be determined in each case.

\acknowledgments

Research supported in part by CNPq, CAPES, FAPERJ and FUJB.

%\end{multicols}
%\end{document}

% &&&&&&&&&&&&&&&&&&&&&&&&&&&&&&&&&&&&&&&&&&&&&&&&&&&&&&&&&&&&&&&&&&&&

%\pagebreak

% ###########################################################
% ############################################################
% ############################################################
% ##############################################################
% ######################################################################
% ##########################################################################
% ##########################################################################
% ##########################################################################
% ##########################################################################

\vspace{4cm}
% ##########################################################################
% ##########################################################################
% #############################################################

\end{multicols}

\end{document}